\documentclass{emulateapj}
\usepackage{lscape}
\usepackage{epsfig}

\slugcomment{Accepted to the Astronomical Journal}

\shortauthors{Bochanski et al.}
\shorttitle{Milky Way M Dwarfs}
\begin{document}

\title{Exploring the Local Milky Way:  M Dwarfs as Tracers of Galactic Populations}

\author{John J. Bochanski\altaffilmark{1},
	 Jeffrey A. Munn\altaffilmark{2},
	 Suzanne L. Hawley\altaffilmark{1},
	 Andrew A. West\altaffilmark{3},
	 Kevin R. Covey\altaffilmark{4},
	 Donald P. Schneider\altaffilmark{5}
	 }

\altaffiltext{1}{Astronomy Department, University of Washington,
   Box 351580, Seattle, WA  98195\\
email: bochansk@astro.washington.edu}
\altaffiltext{2}{US Naval Observatory, Flagstaff Station,
10391 W. Naval Observatory Road, Flagstaff, AZ  86001-8521}
\altaffiltext{3}{Astronomy Department, University of California,
  601 Campbell Hall, Berkeley, CA 94720-3411}
\altaffiltext{4}{Harvard-Smithsonian Center for Astrophysics,
60 Garden Street, MS-72 Cambridge, MA 02138}
\altaffiltext{5}{Department of Astronomy and Astrophysics,
 The Pennsylvania State University, 525 Davey Lab, University Park, PA 16802} 

\begin{abstract}
We have assembled a spectroscopic sample of low-mass dwarfs observed as part of the Sloan Digital Sky Survey along one Galactic sightline, designed to investigate the observable properties of the thin and thick disks.  This sample of $\sim$ 7400 K and M stars also has measured $ugriz$ photometry, proper motions, and radial velocities.  We have computed $UVW$ space motion distributions, and investigate their structure with respect to vertical distance from the Galactic Plane.  We place constraints on the velocity dispersions of the thin and thick disks, using two-component Gaussian fits.  We also compare these kinematic distributions to a leading Galactic model. Finally, we investigate other possible observable differences between the thin and thick disks, such as color, active fraction and metallicity.
\end{abstract}


\section{Introduction}

Modeling the Galaxy presents a challenging breadth of problems to both theorists and observers.  Large N-Body simulations employing the constraints of gravity and $\Lambda$ cold dark matter cosmology have sought to recreate the infant Galaxy, tracing the formation and collapse of baryons within the dark matter halo \citep{2004ApJ...612..894B,2007MNRAS.374.1479G}.  Observationally, these simulations are constrained by rotational velocities and luminosity profiles of extragalactic systems \citep{1997ApJ...482..659D}.
Closer to home, observers seek to reconstruct the history of the Milky Way's cannibalistic mergers through photometric identification of tidal debris, such as the Sagittarius dwarf \citep{1994Natur.370..194I}. Spectroscopy is also employed to find surviving, co-moving stars of similar metallicity (e.g., \citealp{2003ApJ...588..824Y}).
  
A particularly interesting problem to both the theorist and observer is the formation and nature of the thick disk \citep{1983MNRAS.202.1025G}.  This population has been explored extensively, mainly through star counts (e.g. \citealp{1983MNRAS.202.1025G, 1993ApJ...409..635R,1999A&A...348...98B,1999Ap&SS.265..213N,2002ApJ...578..151S}).  However, the thick disk scale height and local density normalization are still uncertain \citep{1999Ap&SS.265..213N}, with values of $h_{\rm thick}$ ranging from $\sim$ 700 - 1500 pc and local normalizations between 2\% and 15\%.  Larger scale heights are usually coupled to lower normalization values (see Figure 1 of \citealp{2002ApJ...578..151S}).  The scale length of the thick disk is also uncertain, though typically scale lengths larger than the thin disk are inferred  \citep{2001ApJ...553..184C,2003AJ....125.1958L}.  The thick disk is thought to be an older, metal-poor population \citep{1993ApJ...409..635R,2000AJ....119.2843C}. Metallicity differences, such as $\alpha$ element enhancement, have been explored \citep{2003A&A...410..527B}, but other observables, such as photometric color and chromospheric activity trends, have yet to be studied in detail.  Thick disks observed in external galaxies \citep{2006AJ....131..226Y} often appear to have structural parameters and kinematics similar to the Milky Way.

Modeling the observable properties of the thin and thick disks, namely star counts \citep{1993ApJ...409..635R, 1993ASPC...49...37R} and kinematics \citep{1996AJ....112..655M,2003A&A...409..523R,2006A&A...451..125V}, has undergone a resurgence in recent years.  The seminal work of \cite{1980ApJS...44...73B} laid the foundation for modeling star counts of the smooth Galactic components.  Present-day models, such as the Besan\c{c}on model \citep{2003A&A...409..523R}, and the Padova model \citep{2006A&A...451..125V} also incorporate kinematics, allowing for robust comparisons to observations.  It is important to rigorously test their predictions against actual observed spatial and kinematic distributions.  

Low-mass dwarfs are both ubiquitous and long-lived \citep{1997ApJ...482..420L}, and serve as excellent tracers of the Galactic potential \citep{1970MNRAS.148..463W, Wielen77, 2006AJ....132.2507W}.  Modern surveys, such as the Sloan Digital Sky Survey (SDSS; \citealp{2000AJ....120.1579Y}), are sensitive to K and early M dwarfs at distances of $\sim$ 1 kpc above the Galactic Plane, probing the transition between the thin \citep{1997PASP..109..559R} and thick \citep{2001A&A...365..424K} disks.  At these distances, we expect about 20\% of the observed stars to be thick disk members, assuming a local normalization of 2\% \citep{1993ApJ...409..635R} and scale heights of 300 pc and 1400 pc for the thin and thick disks, respectively.
Star counts of low-mass dwarfs have been used to determine Galactic structural properties (\citealp{1997PASP..109..559R} and references therein).  These studies sought to determine the vertical scale height and local normalization of the thin and thick disks  \citep{2002ApJ...578..151S}, as well as the underlying mass function of these populations \citep{1998AJ....116.2513M,2000A&A...356..108P,covey_thesis}.  The chemical evolution of the Galaxy has also been explored with red dwarfs \citep{1997PASP..109..559R}, using molecular band indices as a proxy for metallicity \citep{1997AJ....113..806G}.

In addition to their utility as Galactic tracers, low-mass dwarfs have intrinsic properties, such as chromospheric activity \citep{2004AJ....128..426W,2005AJ....130.1871B,2007AJ....133.2258S}, that can be placed in a Galactic context.  Additionally, metallicity may be probed using subdwarfs \citep{2003AJ....125.1598L}, readily identified by their spectra, which show enhanced calcium hydride (CaH) absorption.  Subdwarfs have been easily detected in large-scale surveys such as the SDSS \citep{2004AJ....128..426W}.  

Kinematic studies of low-mass stars have a rich historical background.  Samples are typically drawn from proper motion surveys, such as the New Luyten Two Tenths (NLTT) catalog \citep{1979lccs.book.....L} and the Lowell Proper Motion Survey \citep{1971lpms.book.....G}.  Efforts have been made to identify nearby stars in these surveys, for example by \cite{1991adc..rept.....G}.  However, proper motion surveys possess inherent kinematic bias.  The McCormick sample \citep{Vyssotsky56}, assembled from 875 spectroscopically confirmed K and M dwarfs, has been frequently studied as a kinematically unbiased sample \citep{Wielen77, Weis95, Ratnatunga97}, although it has been suggested that the sample may be biased towards higher space motions \citep{1995AJ....110.1838R}.  The Palomar-Michigan State University survey (PMSU; \citealp{1995AJ....110.1838R, 1996AJ....112.2799H,2002AJ....123.3356G,2002AJ....124.2721R}) is among the largest prior spectroscopic surveys of low-mass stars, obtaining spectral types and radial velocities of $\sim 1700$ M dwarfs.  The PMSU sample, which targeted objects from the Third Catalogue of Nearby Stars \citep{1991adc..rept.....G}, was used to construct a volume complete, kinematically unbiased catalog of $\sim 500$ stars, sampling distances to $\sim 25$ pc.   Other surveys, such as the 100 pc survey \citep{2005AJ....130.1871B}, have also used low-mass stars as kinematic probes.  In Table \ref{table:previous}, we summarize the sample sizes and approximate distance limits of previous major kinematic surveys of low-mass dwarfs, along with the mean velocity dispersions for each study.  It is clear that our sample of several thousand stars out to distances of $\sim 1$ kpc results in a study of Galactic kinematics using low-mass stars with unprecedented statistical significance.

In this paper, we present our examination of the properties of the thin and thick disks using a SDSS Low-Mass Spectroscopic Sample (SLoMaSS) of K and M dwarfs that is an order of magnitude larger than previous samples.  In \S \ref{sec:observations}, we describe the SDSS spectroscopic and photometric observations that comprise SLoMaSS.  The resulting distances and stellar velocities are presented in \S \ref{sec:analysis}.  In \S \ref{sec:results}, we detail our efforts to separate the observed stars into two populations, search for kinematic, metallicity and color gradients and compare our results to a contemporary Galactic model.  Finally, \S \ref{sec:conclusions} summarizes our findings.

\begin{deluxetable*}{llrrrrl}
\tabletypesize{\scriptsize}
\tablecaption{Previous Low-Mass Stellar Kinematic Survey Results}
\tablehead{
\colhead{Study}&
\colhead{Sample}&
\colhead{Distance}&
\colhead{$\sigma_{U}$ \tablenotemark{a}}&
\colhead{$\sigma_{V}$}&
\colhead{$\sigma_{W}$}& 
\colhead{Comments}\\
\colhead{}&
\colhead{Size}&
\colhead{Limit (pc)}&
\colhead{}&
\colhead{}&
\colhead{}&
\colhead{}
}
\startdata

\smallskip\cite{Vyssotsky56, 1995AJ....110.1838R} & 368 & $\sim$ 300 & 39 & 26 & 23  & McCormick stars in PMSU survey\\

\smallskip\cite{Wielen77} & 516 & $\sim$ 300  & 39 & 23 &  20 & McCormick stars\\ 

\smallskip\cite{1995AJ....110.1838R} & 514 & 25  & 43 & 31 & 25 & PMSU I volume-complete sample\\

\smallskip\cite{Ratnatunga97} & 773 & $\sim$ 300  & 30.6 & 18.5 & 7.4 & McCormick stars\\ 

\cite{2002AJ....124.2721R} & 436 & 25  & 37.9 & 26.1 & 20.5 & PMSU IV volume-complete sample \\
\smallskip   					& 	&  25   &  34  & 18    & 16     & PMSU IV volume-complete sample, core\\
\cite{2005AJ....130.1871B} & 419 & 100  & 35 & 21 & 22 & Non-active stars, core\\
\smallskip                                         & 155 & 100  & 19 & 15 & 16 & Active stars, core\\

This study                         & 70 & 100    & 28.4 & 21.3 & 19.2 & $|z|$ $<$ 100 pc\\
					 &    & 100    & 25.7 & 20.9 & 14.1 & $|z|$ $<$ 100 pc, core\\
					 & 4300 & 500    & 39.0 & 30.0 & 24.8 & $|z|$ $<$ 500 pc\\
					 &    & 500    & 34.1 & 24.6 & 20.9 & $|z|$ $<$ 500 pc, core\\ 
					 & 6893 & 1000    & 44.1 & 36.9 & 27.9 & $|z|$ $<$ 1000 pc\\
					 &    & 1000    & 37.4 & 27.2 & 23.6 & $|z|$ $<$ 1000 pc, core\\ 
					 & 7398 & 1600    & 46.6 & 42.2 & 29.3 & Total Sample\\
					 &    & 1600    & 38.5 & 28.6 & 24.4 & Total Sample, core\\ 

 \enddata
 \label{table:previous}
\tablenotetext{a}{The velocity dispersions in $U$,$V$, and $W$ are measured in km s$^{-1}$.}

\end{deluxetable*}

\section{Observations}\label{sec:observations}

\subsection{SDSS Photometry}
The SDSS \citep{2000AJ....120.1579Y,2002AJ....123..485S,2003AJ....125.1559P, 2004AN....325..583I} is a large ($\sim$10,000 sq. deg.), multi-color ($ugriz$; \citealp{1996AJ....111.1748F,1998AJ....116.3040G,2001AJ....122.2129H,2002AJ....123.2121S,2006AN....327..821T}) photometric and spectroscopic survey centered on the Northern Galactic Cap.  The 2.5m telescope \citep{2006AJ....131.2332G}, located at Apache Point Observatory scans the sky on great circles, as the camera \citep{1998AJ....116.3040G} simultaneously images the sky in five bands to a faint limit of $\sim$ 22.2 in $r$, with a typical uncertainty of $\sim$ 2\% at $r \sim 20$ \citep{2003MmSAI..74..978I,2006ApJS..162...38A}.  The last data release (DR5\footnote{http://www.sdss.org/dr5/}; \citealp{DR5}) comprises 8000 sq. deg. of imaging, yielding photometry of $\sim$ 217 million unique objects, including $\sim$ 85 million stars.  SDSS photometry has enabled  a myriad of studies on both Galactic structure (e.g. \citealp{2000ApJ...540..825Y,2002ApJ...569..245N,2005astro.ph.10520J,2006ApJ...642L.137B}) and low-mass stars (e.g. \citealp{2002AJ....123.3409H, 2004PASP..116.1105W, 2005PASP..117..706W,2006PASP..118.1679D}).

\subsection{SDSS Spectroscopy}\label{sec:sloanspec}
When sky conditions prohibit photometric observations, the SDSS telescope is fitted with twin fiber-fed spectrographs.  These instruments simultaneously obtain 640 medium-resolution ($R \sim$ 1,800), flux-calibrated, optical (3800--9200 \AA) spectra per 3$^{\circ}$ plate, permitting radial velocity measurements for most stars with an uncertainty of $\sim$ 10 km s$^{-1}$ \citep{2004AJ....128..502A}.  A typical 45 minute observation yields a signal-to-noise 
ratio per pixel $>$ 4 at $g = 20.2$ and $i = 19.9$ \citep{2002AJ....123..485S}, with a broadband flux calibration uncertainty of  $\sim 4\%$ \citep{2004AJ....128..502A}. The DR5 sample includes over 1 million spectra, with $\sim$ 216,000 stellar spectra \citep{DR5}.   The majority of spectra in the SDSS database are drawn from 3 main samples, which target objects based on their photometric colors and morphological properties.  These samples are optimized to observe galaxies \citep{2002AJ....124.1810S}, luminous red galaxies with z $\sim 0.5 - 1.0$ \citep{2001AJ....122.2267E}, and high redshift quasars \citep{2002AJ....123.2945R}.  However, low-mass stars have similar colors to some of these samples and therefore are observed serendipitously.
SDSS spectroscopy of late-type dwarfs has been the focus of numerous studies (see \citealp{2002AJ....123.3409H,2003AJ....125.2621R,2004AJ....128..426W,2006AJ....131.1674S,2006AJ....132.2507W,2007AJ....133..531B} and references therein). 

\subsection{SDSS Low-Mass Spectroscopic Sample: SLoMaSS}
During Fall 2001, a call was placed to the SDSS collaboration to design special spectroscopic plates that employed different targeting algorithms than the usual SDSS survey samples described in \S \ref{sec:sloanspec}.  We designed and proposed a series of observations to probe the local vertical structure of the Milky Way, obtaining spectra of low-mass dwarfs in the southern equatorial stripe (stripe 82) of the SDSS photometric footprint during Fall 2002.  This stripe at zero declination is repeatedly observed during the time of the year when the Northern Galactic Cap is not visible.  These repeat scans sample an area of $\sim 300$ sq. deg. and have been used to study Type Ia supernovae \citep{2005astro.ph..4455S, f07}, stellar variability \citep{2007arXiv0704.0655S} and characterize the repeatability of SDSS photometry \citep{2007astro.ph..3157I}. SLoMaSS is comprised of stars with unsaturated $griz$ photometry and extinction-corrected magnitude limits of 15 $< i <$ 18 and $i-z > 0.2$.   A series of three spectroscopic tilings composed of 15 plates (numbers 1118-1132, centered on $l \sim 105^\circ$, $b \sim -62^\circ$) was observed, yielding a total of 8880 stellar spectroscopic targets in SLoMaSS, each with $ugriz$ photometry.  

\section{Analysis}\label{sec:analysis}
\subsection{Spectral Types}
\begin{figure}[htbp]
    \centering
       \includegraphics[scale=0.34, angle=90]{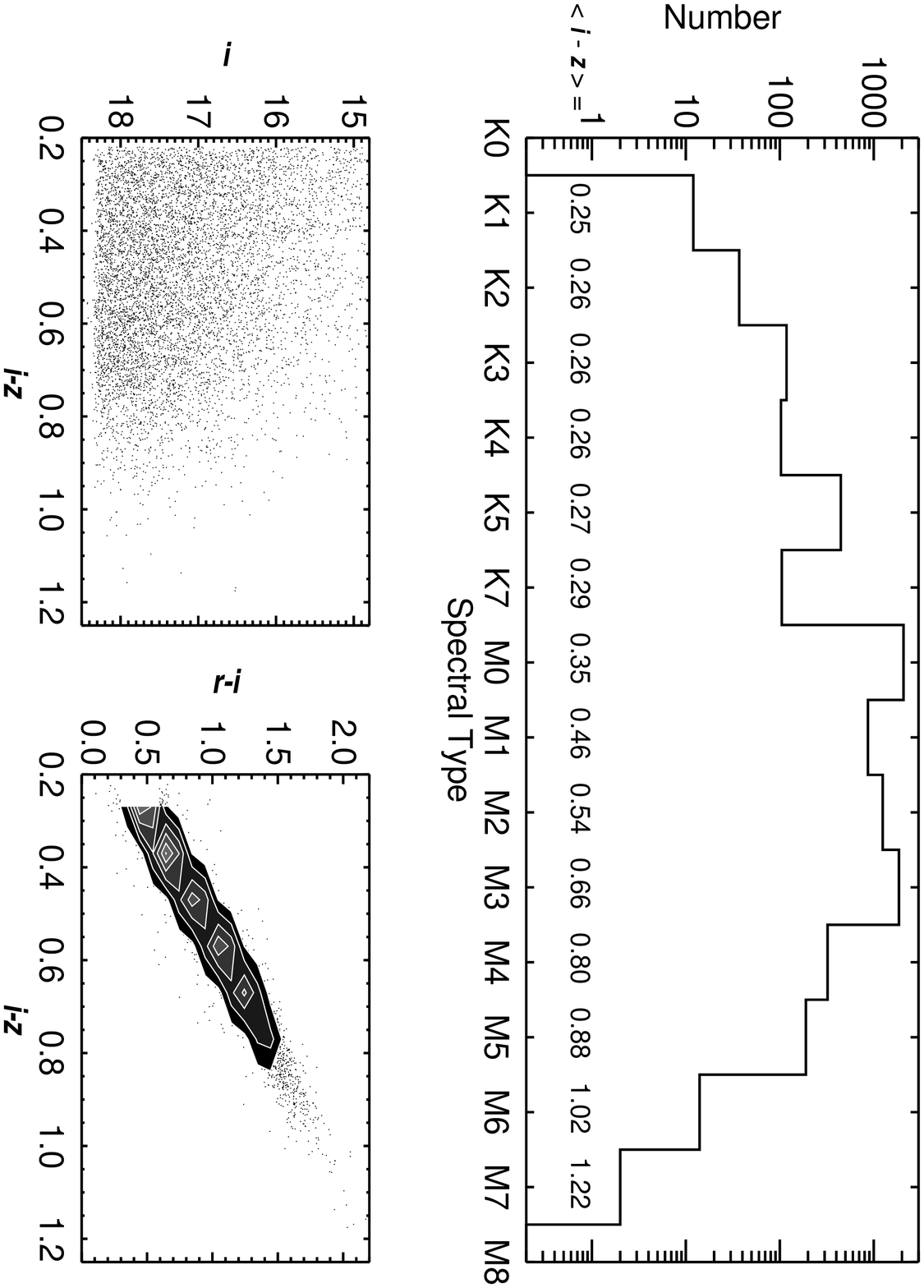}
    \caption{Upper panel: Distribution of spectral types within SLoMaSS.  The mean $i-z$ color at each spectral type is shown above each bin. Lower panels: $i$ vs. $i-z$ color-magnitude distribution and $r-i$ vs. $i-z$ color-color diagram of stars in SLoMaSS.  The lowest contour (black) in the color-color diagram indicates a density of 100 stars in a square 0.1 magnitudes wide.  Each additional contour represents 100 additional stars.}
\label{fig:spectypes}
\end{figure}

The spectroscopic data were analyzed with the HAMMER suite of software \citep{c07} which measures spectral type, H$\alpha$ emission line properties, and various spectral band indices.  This pipeline automatically assigns a spectral type to each star, then allows the user to confirm and adjust the spectral type manually, if necessary.  The accuracy of the final spectral types is $\pm$ 1 subtype.  After examining the entire sample by eye, we selected spectroscopically confirmed K and M dwarfs. This cut decreased the sample from 8880 to 8696 stars.  The final spectral type distribution of SLoMaSS, after the additional cuts explained below, together with the average $i-z$ color for each spectral type bin, is shown in Figure \ref{fig:spectypes}.  

\subsection{Distances - Photometric Parallax}
\begin{figure}[htbp]
    \centering
      \includegraphics[scale=0.35, angle=90]{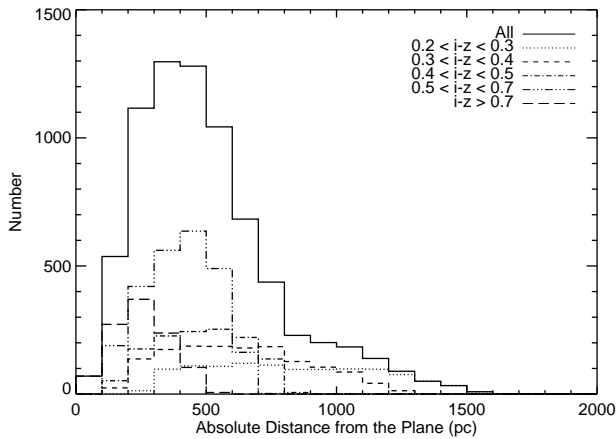}
       \caption{The absolute vertical distance distribution of the sample binned every 100 pc.  The solid thick histogram is the total distribution, while the five remaining histograms represent the height distribution for five $i-z$ color bins as described in the legend.  Note that there are hundreds of stars per bin out to an absolute vertical distance of $\sim$ 1000 pc.}
\label{fig:vert}
\end{figure}

Distances to the SLoMaSS stars were determined using the photometric parallax relations described in \cite{2005PASP..117..706W} and \cite{2006PASP..118.1679D}.  The \cite{2005PASP..117..706W} relation was employed for stars with $i-z$ redder than 0.37 and the \cite{2006PASP..118.1679D} parallax relation was applied to 
the bluer stars in SLoMaSS.  We neglect reddening corrections, since the average extinction computed for the total column along this line of sight from \cite{1998ApJ...500..525S} was small ($< 0.05$ in $i$), and most of these stars will lie in front of significant amounts of dust. Stars with $i-z$ colors that did not fall within the appropriate boundaries of the photometric parallax relations (0.22 $< i-z <$ 1.84) were removed from the sample, decreasing its size to 8280 stars.  Additionally, we applied the photometric white dwarf-M dwarf pair cuts of \cite{2004ApJ...615L.141S}, removing 4 more stars from the sample.  The $i, i-z$ Hess diagram and $r-i, i-z$ color-color diagram are shown in the lower panels of Figure \ref{fig:spectypes}.  
The distance from the Galactic Plane was computed assuming the Sun's vertical position to be 15 pc above the Plane \citep{1995ApJ...444..874C,1997A&A...324...65N,1997MNRAS.288..365B}.  The vertical distribution of stars in SLoMaSS is shown in Figure \ref{fig:vert}.  Note that the decline in stellar density at large distances reflects the incompleteness of our sample, not the intrinsic stellar density, while the decline at small distances from the Plane is due to the saturation of SDSS photometry.

\subsection{Distances - Spectroscopic Parallax}\label{sec:specparallax}
In addition to the photometric parallax relations mentioned above, we also computed the distances to the M dwarfs in SLoMaSS using the spectroscopic parallax relations of \cite{2002AJ....123.3409H}.  These distances served as a check on the photometric parallax relations and were later used to divide the sample in a search for color differences between the thin and thick disks (see \S \ref{sec:colors} for details).  The spectroscopic parallax of K stars was not computed, as no reliable calibrated relations were available.

\subsection{Radial Velocities, Proper Motions \& $UVW$ Velocities}
Radial velocities were computed by cross-correlating the M dwarf stellar spectra against the low-mass template spectra of \cite{2007AJ....133..531B} with the IRAF\footnote{
IRAF is distributed by the National Optical Astronomy Observatories,
which are operated by the Association of Universities for Research
in Astronomy, Inc., under cooperative agreement with the National
Science Foundation.} task $fxcor$.  This cross-correlation routine is based on the method described in \cite{ 1979AJ.....84.1511T}.  The measured velocities have external errors of $\sim$ 4 km s$^{-1}$.  The K dwarf radial velocities were obtained directly from the SDSS/Princeton 1D spectral pipeline\footnote{http://spectro.princeton.edu}, with typical uncertainties of 10 km s$^{-1}$. 

Proper motions were determined from the SDSS+USNO-B proper motion catalog \citep{2004AJ....127.3034M}.  The catalog is $\sim90\%$ complete over the magnitude limits of our sample, with random errors of $\sim$ 3.5 mas yr$^{-1}$.  We removed an additional 878 stars with poorly measured proper motions\footnote{Specifically, we required a match between catalogs (match $>$ 0), detections in at least 4 plates in USNO-B (nfit $\geq$ 5) , no other objects within 7 arcseconds, which is the resolution of the POSS plates (dist22 $>$ 7; see \citealp{2006AJ....131..582K}), and small errors in the proper motion determination (sigRA $<$ 1000 and sigDec $<$ 1000); see \citep{2004AJ....127.3034M} for details.}, reducing our final sample size to 7398 stars.

Combining distances, proper motions and radial velocities, we computed the $UVW$ space motions of each star in SLoMaSS using the method of \cite{1987AJ.....93..864J}.  The velocities are computed in a right-handed coordinate system, with positive $U$ velocity directed toward the Galactic center and corrected for the solar motion (10, 5, 7 km s$^{-1}$; \citealp{Dehnen98}) with respect to the local standard of rest. 




\section{Results}\label{sec:results}
The SLoMaSS observations were used to investigate kinematics, colors, chromospheric activity and metallicity in the thin and thick disk populations.

We used two methods in our analysis.  First, we examined the velocity distributions with no assumptions regarding the parent population of a given star (see \S \ref{sec:kinematics}).  These kinematic results were then used to test a leading Galactic model (\S \ref{sec:model}).  We also used the space motion of each star to assign it to the thin or thick disk population, and investigated the color, metallicity and activity differences between the two populations (\S \ref{sec:diffs}).

\subsection{Kinematics}\label{sec:kinematics}
\begin{figure}[htbp]
    \centering
       \includegraphics[scale=0.35, angle=90]{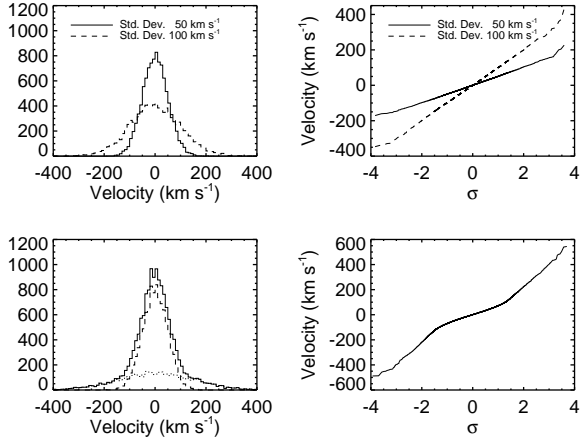}
    \caption{Schematics of velocity distributions (left panels) and their corresponding probability plots (right panels).  The upper row displays two Gaussian distributions, one with a standard deviation of $\sim 50$ km s$^{-1}$ (solid line) and another with a standard deviation of $\sim 100$ km s$^{-1}$ (dashed line).  A straight line in the probability plot signifies an underlying Gaussian distribution and the slope of the line is a measure of the standard deviation of the distribution.  Note that the slope of the dashed line is twice that of the solid line.  In the bottom row, a low dispersion component (dashed line) and high dispersion component (dotted line) are summed to produce the solid histogram.  While the dual nature of the distribution is not readily apparent in the solid histogram, it is easily detected in the probability plot.}
\label{fig:probplots}
\end{figure}

The sample was binned in 100 pc increments of vertical distance from the Galactic Plane and probability plots of $UVW$ velocities were constructed \citep{1980AJ.....85.1390L, 1995AJ....110.1838R, 2002AJ....124.2721R}.  These diagrams plot the cumulative probability distribution in units of the standard deviation of the distribution.  Hence, a single Gaussian distribution will appear as a straight line with a slope corresponding to the standard deviation of the sample and the y-intercept equal to the median of the distribution. Non-Gaussian distributions will significantly deviate from a straight line.  An example of velocity distributions and their probability plots are shown in Figure \ref{fig:probplots}. An advantage to this method is its immunity to outliers and to binning effects from poorly populated histograms.

\begin{figure}[htbp]
    \centering
       \includegraphics[scale=0.35, angle=90]{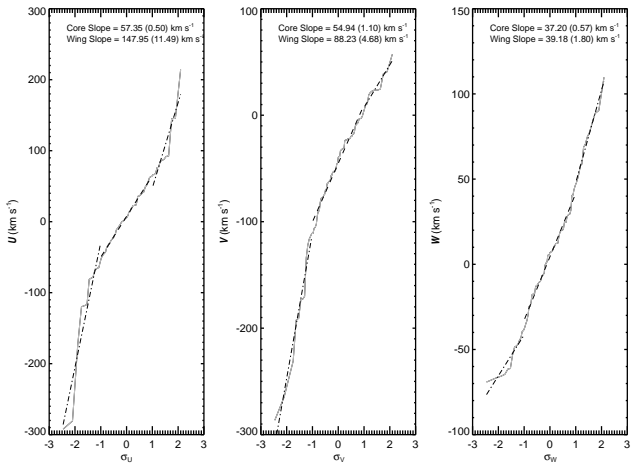}
    \caption{Illustrative example of our velocity dispersion analysis.  Shown are probability plots for $U,V$ and $W$ for the 1200 $<$ $z$ $<$ 1300 pc bin (shaded lines).  The ``core'' component (dashed line) is fit between $\pm$ 1 $\sigma$, while the ``wing'' component (dot-dashed line) fits the outer edges of the distribution.  Note the strong asymmetric drift component manifested as a change in slope at negative $V$ velocities.}
\label{fig:uvwcdf}
\end{figure}

For each distance bin in SLoMaSS, the probability plots are well fit with two lines: a low-dispersion, kinematically colder ``core'' component  ($<~|~1\sigma|$) and a high-dispersion, ``wing" component ($>~|~1\sigma|$).  Shown in Figure \ref{fig:uvwcdf} is an example of our analysis, illustrating linear fits to the core and wing distributions.  At larger Galactic heights, the wing component traces the \textit{in situ} thick disk population.  

\begin{figure}[htbp]
    \centering
       \includegraphics[scale=0.35, angle=90]{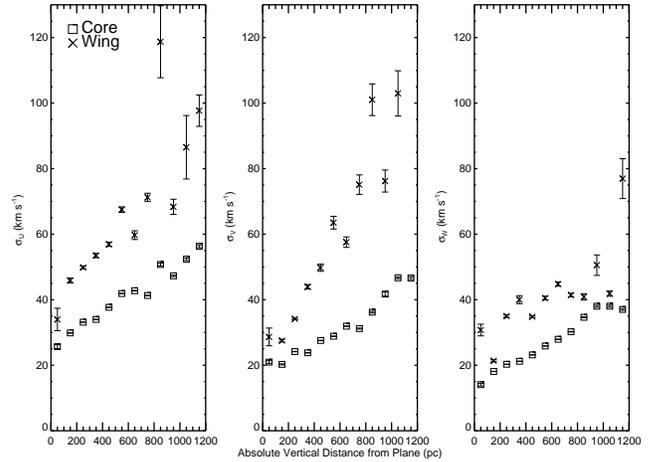}
    \caption{Velocity Dispersion as a function of vertical distance from the plane in 100 pc bins.  The open squares represent the kinematically colder, core component and the crosses represent the high-dispersion, wing component.  Errors are derived from the linear fits to the probability plots.  Note the smooth trend in the low-dispersion component as a function of height from the Plane.}
\label{fig:veldisps}
\end{figure}

The resulting dispersions are shown as a function of distance from the Galactic Plane in Figure \ref{fig:veldisps}.  The low-dispersion core component is well-behaved, smoothly increasing with increasing height.  The high-dispersion wing component is subject to larger scatter, but generally increases with height from the Galactic Plane.  Our results are summarized in Table \ref{table:previous}.  In order to facilitate comparison to the previous results included in the Table, we report the dispersions for several distance bins, as well as the entire sample.


\subsection{Galactic Dynamical Models}\label{sec:model}
\subsubsection{Besan\c{c}on Model: Introduction}
Our sample is well suited to rigorously test current Galactic models.  We chose the Besan\c{c}on model \citep{2003A&A...409..523R} as a fiducial, since it simulates both star counts and kinematics.  This model is constructed on several assumptions and empirical constraints, with the goal of reproducing the stellar content of the Milky Way.  Four populations comprise the model: the thin and thick disks, bulge and halo.  For each population, a star-formation history, initial mass function, density law and age are imposed.  Additionally, an age-metallicity relation is employed for each population, with a Gaussian dispersion about the mean metallicity of each component (see Tables 1 and 3 of \citealp{2003A&A...409..523R}).  

Kinematically, the thin disk is composed of 7 groups of different ages (from 0.0 to 10 Gyr), each being isothermal except for the youngest (0.0 - 0.15 Gyr).  The thick disk is composed of an 11 Gyr old population, with a velocity ellipsoid based on the measurements of \cite{1996A&A...311..456O,1999A&A...351..945O}.  An age-velocity dispersion relation from \cite{1997hipp.conf..621G} is imposed on each component, and the model is then allowed to self-consistently evolve to the present-day.  This self-consistency is achieved with the method described in \cite{1987A&A...180...94B}.  Stellar populations are formed according to their appropriate density laws and introduced with an initial velocity dispersion.  The mass density of the stars is summed in a column of unit volume centered on the Sun along with dark matter halo and interstellar material contributions, and the potential is computed using the Poisson equation.  Stars are then evolved using the collisionless Boltzmann equation in this new potential, and redistributed in the $z$ direction, as complete orbital evolution is not included in this model.  The process is iterated until the potential and scale heights of the disk populations converge at the 1\% level.  The local-mass density, which was empirically determined by \cite{1998A&A...329..920C}, also imposes a constraint on the initial Galactic potential.  We note that imposing an age-velocity dispersion relation on the model dominates the kinematics, and that there are significant uncertainties in these relations.  For example, the age-velocity dispersions used by \cite{2004A&A...423..517R} differ by factors of $\sim$ 2 for the oldest stars, compared to the relations in \cite{1997hipp.conf..621G}.  These discrepancies are primarily derived from the difficulty in determining ages of field stars, with systematic differences imposed by various methods (i.e. isochrone fitting vs. chromospheric ages).

Intrinsic properties such as age, mass, luminosity, metallicity, position and velocity are available for each star in the simulation.  Observable properties, such as colors, proper motions and radial velocities are also reported.  To determine colors, the model uses the \cite{1998A&AS..130...65L} database, which employs adjusted stellar synthetic spectra that attempt to match empirical color-temperature relations.

\subsubsection{Comparison to SDSS}
We queried the Besan\c{c}on webpage\footnote{http://physique.obs-besancon.fr/modele/}, generating a suite of synthetic datasets along the appropriate Galactic sightline of SLoMaSS ($l \sim 105^\circ$, $b \sim -62^\circ$) with proper magnitude limits and error characteristics.  We included both the thin and thick disks in the model inputs.  The bulge and halo components were excluded since SLoMaSS points away from the bulge, and with a maximum distance of $\sim$ 2000 pc, we expect halo contamination to be minimal.  A total of 25 models were generated from identical input, in order to minimize Poisson noise.

As a consistency test of the Besan\c{c}on model, we compared the model star counts to those obtained with SDSS survey photometry in the SLoMaSS field.  We queried the SDSS Catalog Archive Server\footnote{http://cas.sdss.org/dr5/en/} for good point spread photometry\footnote{Specifically, we required the following flags:  a detection in BINNED1 and no EDGE, NOPROFILE, PEAKCENTER, NOTCHECKED, PSF\_FLUX\_INTERP, SATURATED, BAD\_COUNTS\_ERROR, DEBLEND\_NOPEAK, CHILD, BLENDED, INTERP\_CENTER or COSMIC\_RAY flags set \citep{2002AJ....123..485S}.}  of stars along the appropriate sightline.    This query resulted in 46,730 stellar targets.  In the models, the mean star count was 46,991 with a standard deviation of 137.  This excellent agreement should be viewed cautiously, as only rough magnitude cuts were imposed on the Besan\c{c}on models, and we did not attempt to model observational problems that would affect SDSS star counts \citep{2005AJ....129.2047V}, such as cosmic rays or diffraction spikes near bright stars.  However, the test inspires confidence that the Besan\c{c}on model adequately represents the observed SDSS star counts.  

\subsubsection{Comparison to SLoMaSS}
\begin{figure}[htbp]
   \centering
      \includegraphics[scale=0.35, angle=90]{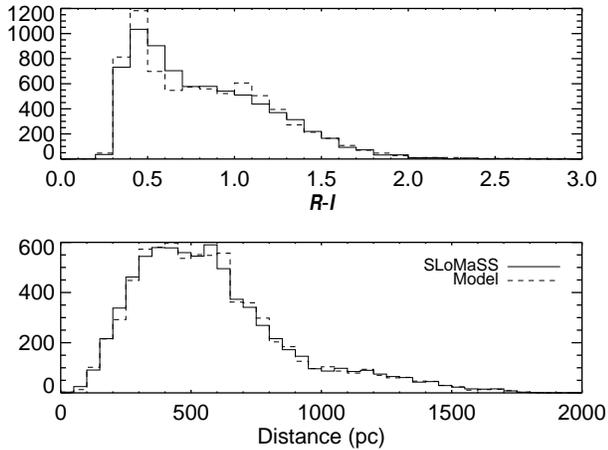}
      \caption{Shown are the $R-I$ and distance distributions of the stars in SLoMaSS (solid line) and a sampled model (dashed line).  Note the agreement between the two datasets, indicating that we are sampling each instance of the Besan\c{c}on model in a manner consistent with the spectroscopic observations in SLoMaSS.}
\label{fig:modelcomp}
\end{figure}

Each of the 25 models was sub-sampled to reproduce the color and distance distributions observed in SLoMaSS.  
The two-dimensional color-distance density distribution was computed for SLoMaSS, and color-distance pairs were randomly drawn according to this normalized density distribution.  If the color-distance pair was found in the Besan\c{c}on model, then it was kept.  This Monte-Carlo sampling continued until there were 7398 color-distance pairs, matching the number of stars in SLoMaSS.
This sampling forced each model to simulate the properties of the SLoMaSS observations when compared to the complete SDSS photometry .  Since SDSS $r-i$ colors are not available for the Besan\c{c}on model, the SLoMaSS colors were transformed to $R-I$ using the relations of \cite{2006PASP..118.1679D}.   Shown in Figure \ref{fig:modelcomp} are the $R-I$ and distance distributions for one instance of the model, along with those from SLoMaSS.  Thus, each model is ``observed" in a manner consistent with the spectroscopic observations of the SDSS field photometry.

\subsubsection{Kinematic Comparisons}\label{sec:modelkins}
Using the sub-sampled model data, we compared the kinematic predictions of the Besan\c{c}on models to the $UVW$ velocities of SLoMaSS.  Using the method described in \S \ref{sec:kinematics}, we constructed two sets of probability plots for each model: one composed of the thin and thick disk stars and one measured solely for the thin disk.  The reason for this separation was twofold.  The first was to isolate any systematics between the models thin and thick disk predictions.  Additionally, this allowed for direct testing comparison of the thin disk predictions, since most of these stars lie on the inner ``core'' region of the probability plot.  That is, if the velocity predictions are correct for the thin disk, the overall slope of the isolated thin disk models should roughly match the ``core'' slope measured from SLoMaSS.

\begin{figure}[htbp]
     \centering
       \includegraphics[scale=0.35, angle=90]{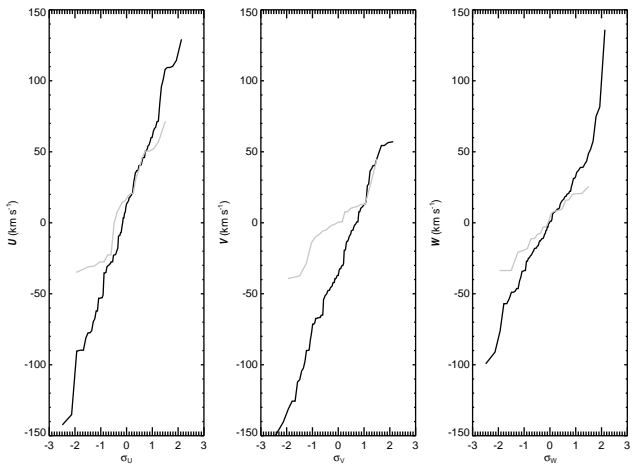}
        \caption{Probability Plots of the $U,V$ and $W$ velocities for pure thin disk (shaded line) and thin + thick models (solid line) in the 1200 $<$ $z$ $<$ 1300 pc bin.  This height bin corresponds to the SLoMaSS data shown in Figure \ref{fig:uvwcdf}.  Note that when the thick disk is added to the distribution, the overall slope increases, and the wing components are enhanced, as seen in the $W$ probability plot.  The combined thin and thick disk velocity distributions are also necessary to match the two-component structure seen in the SLoMaSS probability plots.}
 \label{fig:model_probs}
 \end{figure}

In Figure \ref{fig:model_probs}, an illustrative example of this analysis is shown.  The main effect of adding the thick disk component is to increase the slope of both the core and wing components.  Additionally, the wing component is enhanced, as seen in the $W$ velocity distribution, which is expected from addition of a high dispersion population.  It is clear from comparing to Figure \ref{fig:uvwcdf} that the combination of the thin and thick disk model predictions are necessary to simulate the structure seen in the SLoMaSS data.

\begin{figure*}[htbp]
    \centering
      \includegraphics[scale=0.55, angle=90]{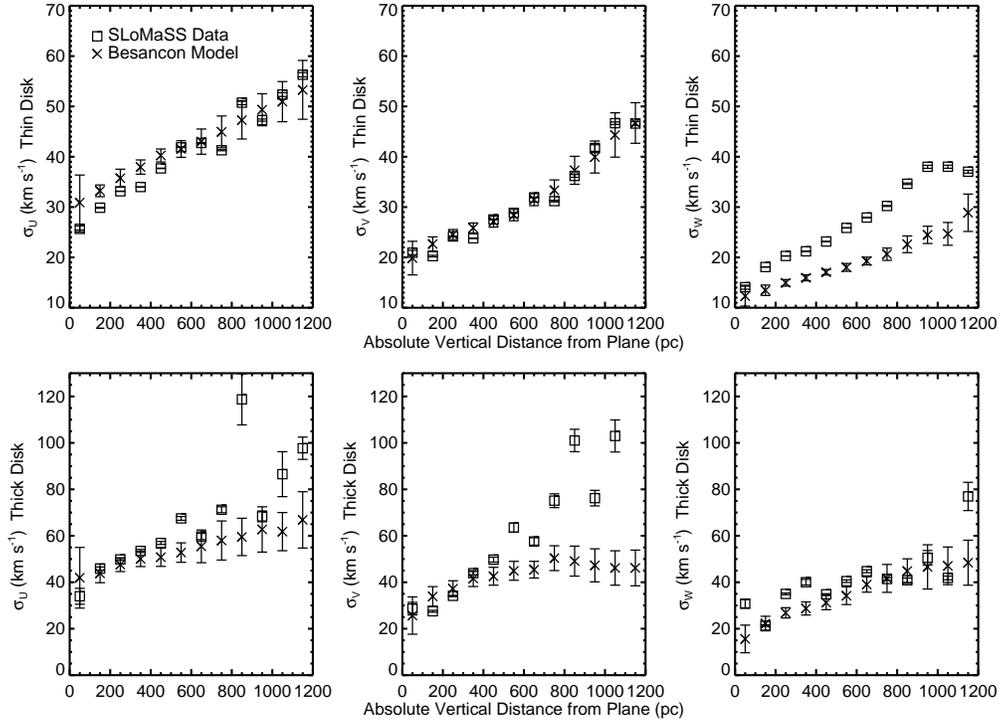}
       \caption{Thin Disk (upper panels) and Thick Disk (lower panels) velocity dispersions for SLoMaSS data (open squares) and Besan\c{c}on model predictions (crosses).  The model reproduces general trends, such as increased dispersion with distance from the Galactic Plane.  The thin disk $W$ velocity dispersions are systematically low, as are the thick disk $V$ dispersions at large heights.  This is probably due to the assumed age-velocity dispersion relation, as explained in \S \ref{sec:modelkins}.}
\label{fig:disp_comp}
\end{figure*}

\begin{figure*}[htbp]
    \centering
      \includegraphics[scale=0.55, angle=90]{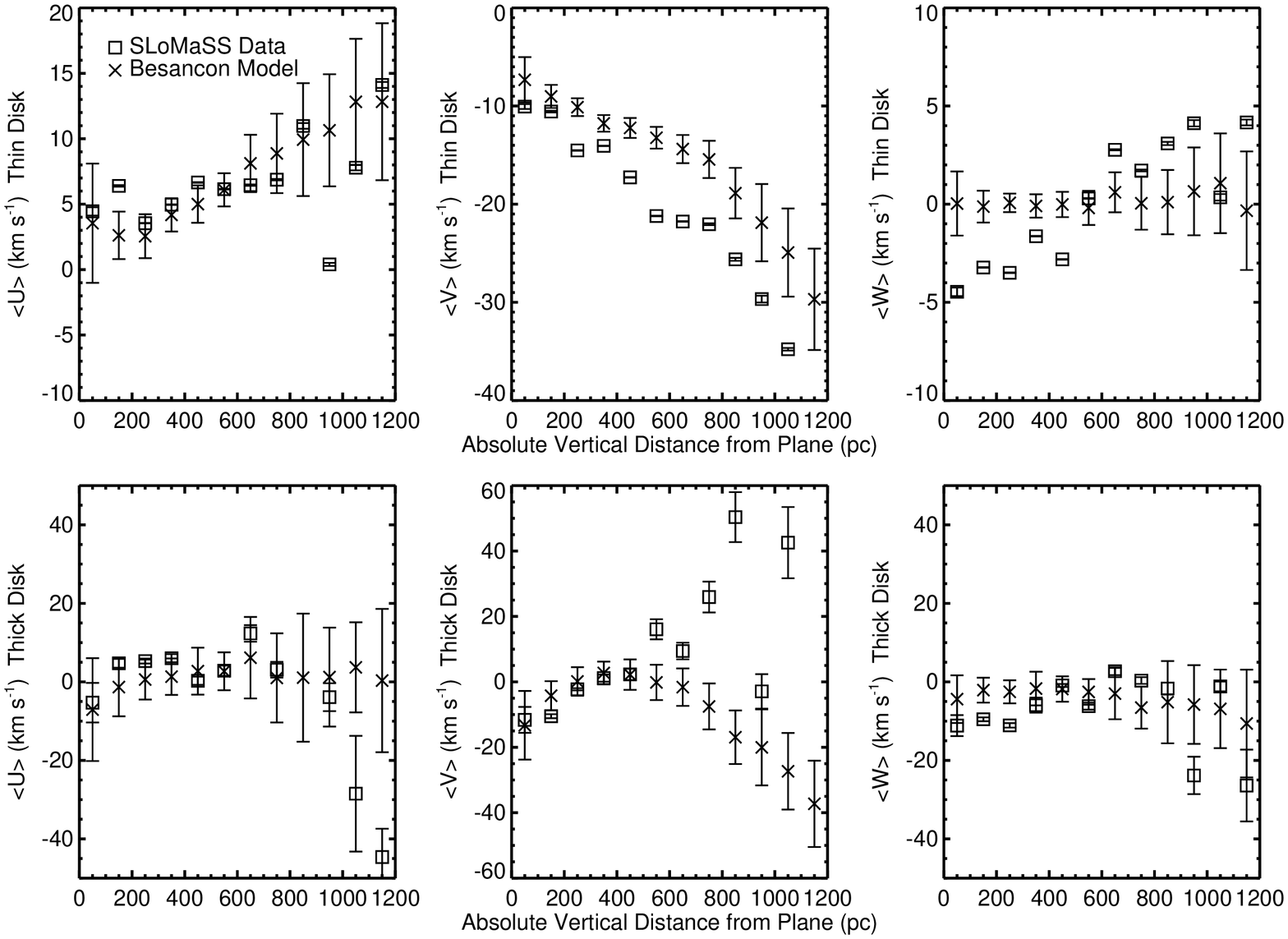}
       \caption{Thin Disk (upper panels) and Thick Disk (lower panels) mean velocities for SLoMaSS data (open squares) and Besan\c{c}on model predictions (crosses). The thin disk mean $V$ velocities predicted by the Besan\c{c}on models are systematically low, while the observed thick disk mean $V$ velocities diverge at large Galactic heights. }
\label{fig:mean_comp}
\end{figure*}

Following the method explained in \S \ref{sec:kinematics} we measured the slopes of the ``core'' and ``wing'' components of the each model as a function of distance from the Galactic Plane.  The results of this analysis, compared to the SLoMaSS results are shown in Figures \ref{fig:disp_comp} and \ref{fig:mean_comp}.  The SLoMaSS results (open squares), which are shown in Figure \ref{fig:veldisps}, are compared to the average Besan\c{c}on prediction (crosses) for the thin (upper panels) and thick disk (lower panels).  
While the model does well in predicting general trends, there are clearly some systematics.   
The model predictions for the $\sigma_W$ thin disk velocity dispersions are systematically low, suggesting there may be flaws in the method used to compute these motions. Additionally, the model underestimates the thick disk $\sigma_V$ dispersions at large Galactic heights.  As described above (\S \ref{sec:model}), the assumed age-velocity dispersions relations, which dominate the predicted kinematic structure, are uncertain and may contribute to this disagreement.  This speculation is supported by the results summarized in Table \ref{table:previous}, which demonstrate that most previous surveys (as well as our own) have measured velocity dispersions significantly higher than those predicted from the model.  The mean velocity distributions for both SLoMaSS and the Besan\c{c}on model are shown in Figure \ref{fig:mean_comp}.  Again, the model in general performs well, but there are evident systematic differences.  The mean $V$ velocities predicted for the thin disk by the model are systematically high, and the thick disk $V$ velocities diverge at large Galactic heights.  The first difference may be attributed to our chosen solar motion values (10, 5, 7 km s$^{-1}$; \citealp{Dehnen98}).  A larger adopted value of $V_\odot$, such as the classic value of 12 km s$^{-1}$ \citep{1965gast.conf...61D}, would move the mean velocities towards agreement.  The discrepancy in the thick disk $V$ velocities is probably due to small number statistics, as seen in Figure \ref{fig:vert}.


\subsection{Differences between the Thin and Thick Disks}\label{sec:diffs}
In order to examine observable differences between M dwarf members of the thin and thick disks, we kinematically separated the sample using the method of \cite{2003A&A...410..527B}.  This technique selects outliers in the wings of the three-dimensional Gaussian velocity distribution, and computes the probability of these stars belonging to the thick disk.  The thin and thick disk populations were characterized by the velocity dispersions in \cite{2003A&A...410..527B}. In order to minimize systematics, the $UVW$ velocities were re-computed using distances determined with the spectroscopic parallax relations of \cite{2002AJ....123.3409H}, as described above in \S \ref{sec:specparallax}.  Thus, $UVW$ should not \textit{a priori} vary systematically with color.  This also limits the analysis to the 6577 M dwarfs in the sample.

\subsubsection{Metallicity}\label{sec:metals}
\begin{figure}[htbp]
    \centering
       \includegraphics[scale=0.55, angle=90]{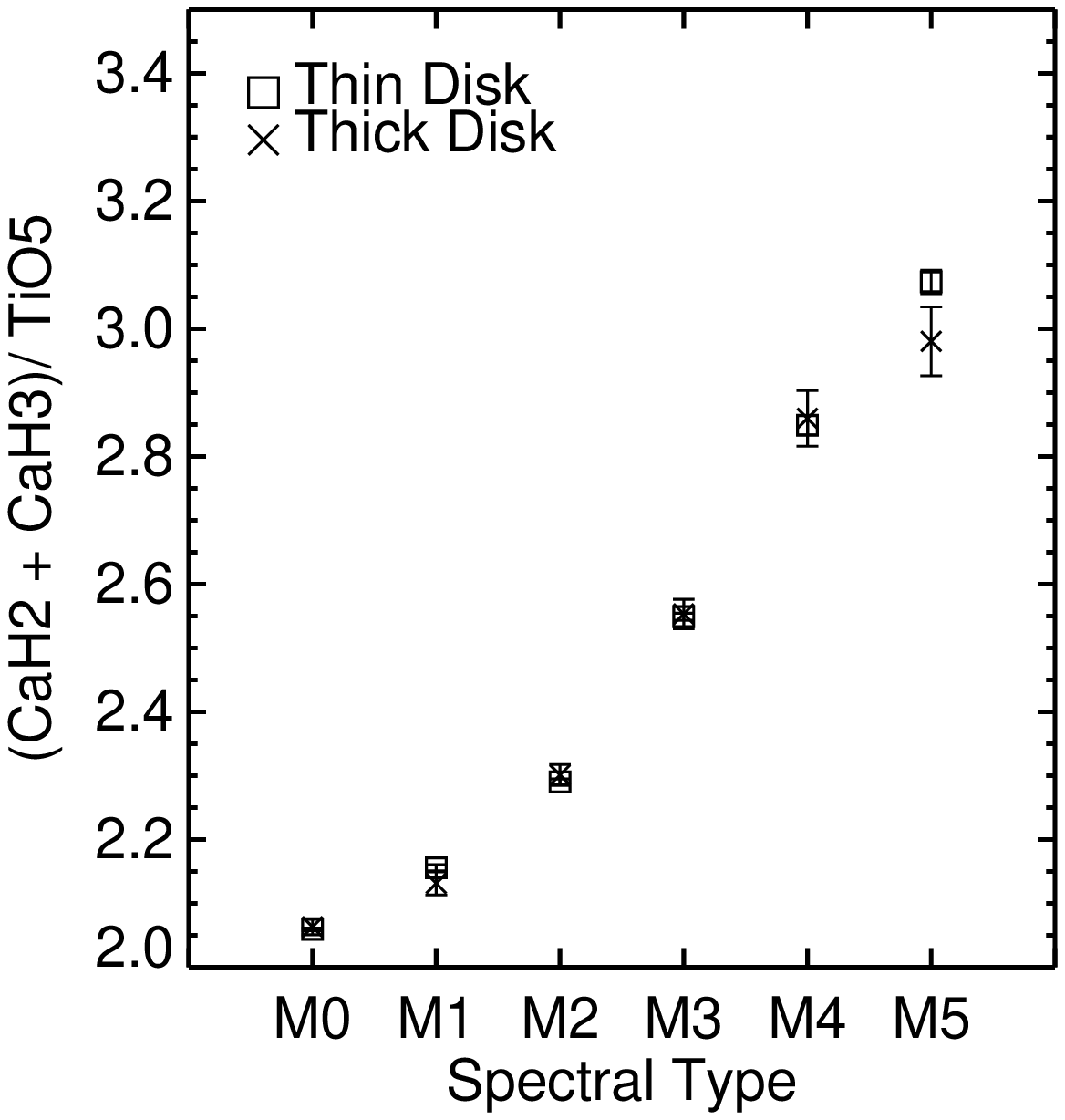}
    \caption{The metallicity-sensitive ratio of (CaH2 + CaH3) / TiO5 vs. Spectral Type for the thin (open squares) and thick (crosses) disk populations.  Higher ratio values indicate a higher metallicity \citep{Burgasser06}.  Within the error bars, the thin and thick disk populations exhibit very similar behavior, indicating that the sample may not be probing a large spread ($\gtrsim$ 1 dex) in metallicity.}
\label{fig:metals}
\end{figure}

The exact formation mechanism of the thick disk is uncertain (see \citealp{1993ARA&A..31..575M} and references therein), but there is evidence that the mean metallicity of the thick disk is lower than that of the thin disk \citep{1993ApJ...409..635R,2000AJ....119.2843C,2003A&A...410..527B}.  Additionally, differences in the $\alpha$-element distributions have been shown to be distinct \citep{1998A&A...338..161F,2003A&A...397L...1F,2003A&A...410..527B}.  
While direct metallicity determinations of M dwarfs are difficult \citep{2006PASP..118..218W}, proxies of metallicity have been employed \citep{1997AJ....113..806G,2003AJ....125.1598L, Burgasser06}.  These previous studies have used the low-resolution molecular band indices (CaH1, CaH2, CaH3, and TiO5) defined in \cite{1995AJ....110.1838R} to roughly discriminate between solar-metallicity, subdwarf ([$m/H$] $\sim$ -1.2) and extreme-subdwarf ([$m/H$] $\sim$ -2) populations.  Adapting the methods of \cite{2003AJ....125.1598L} and \cite{Burgasser06} we computed the ratio (CaH2 + CaH3) / TiO5 for each star in the sample, which varies such that a larger value indicates a higher metallicity.  The mean ratio for each spectral type for the thin and thick disk populations is shown in Figure \ref{fig:metals}.  The two populations do not separate within the errors, suggesting that the observed metallicity distributions do not differ greatly.

\subsubsection{Color Gradients}\label{sec:colors}
\begin{figure}[htbp]
    \centering
       \includegraphics[scale=0.55, angle=90]{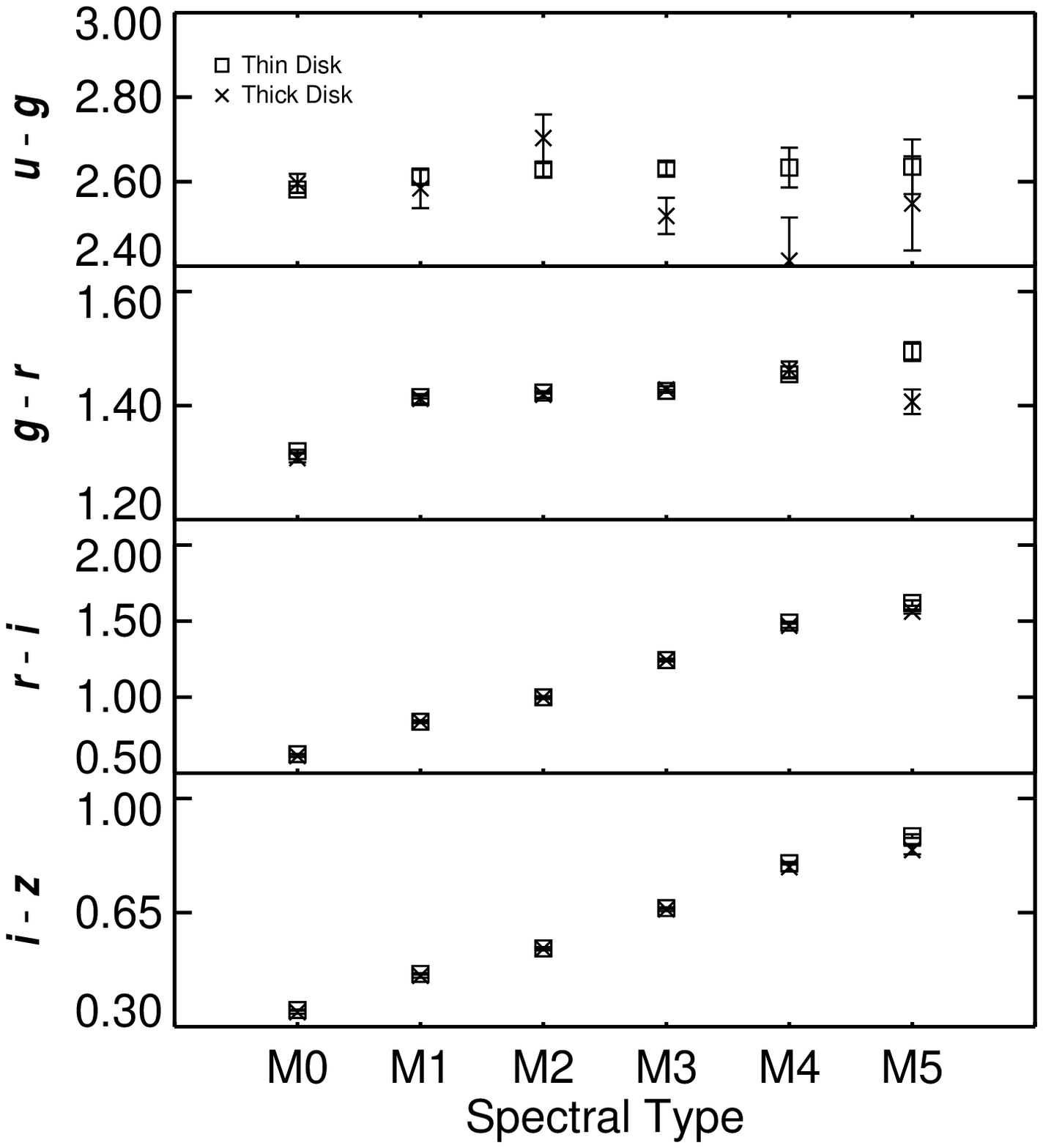}
    \caption{The $ugriz$ colors as a function of spectral type for the thin (open squares) and thick (crosses) disk populations.  The colors of both populations agree within the errors, suggesting that the M dwarfs in SLoMaSS do not cover a large metallicity range.}
\label{fig:colors}
\end{figure}

A lower mean metallicity could also manifest itself as a color shift at a given spectral type. Specifically, \cite{2004AJ....128..426W} showed that at a given $r-i$ or $i-z$ color, low metallicity subdwarfs ([$m/H$] $\sim -1.2$) were redder in $g-r$.  
After SLoMaSS was kinematically separated, the mean $u-g, g-r, r-i$ and $i-z$ colors of the thin and thick disk stars were computed at each spectral subtype, as shown in Figure \ref{fig:colors}.  There are no significant color differences between the thin and thick disks at a given spectral type.  This further suggests that the SLoMaSS stars are not probing a large spread in metallicity.

\subsubsection{Chromospheric Activity}
\begin{figure}[htbp]
    \centering
       \includegraphics[scale=0.45, angle=90]{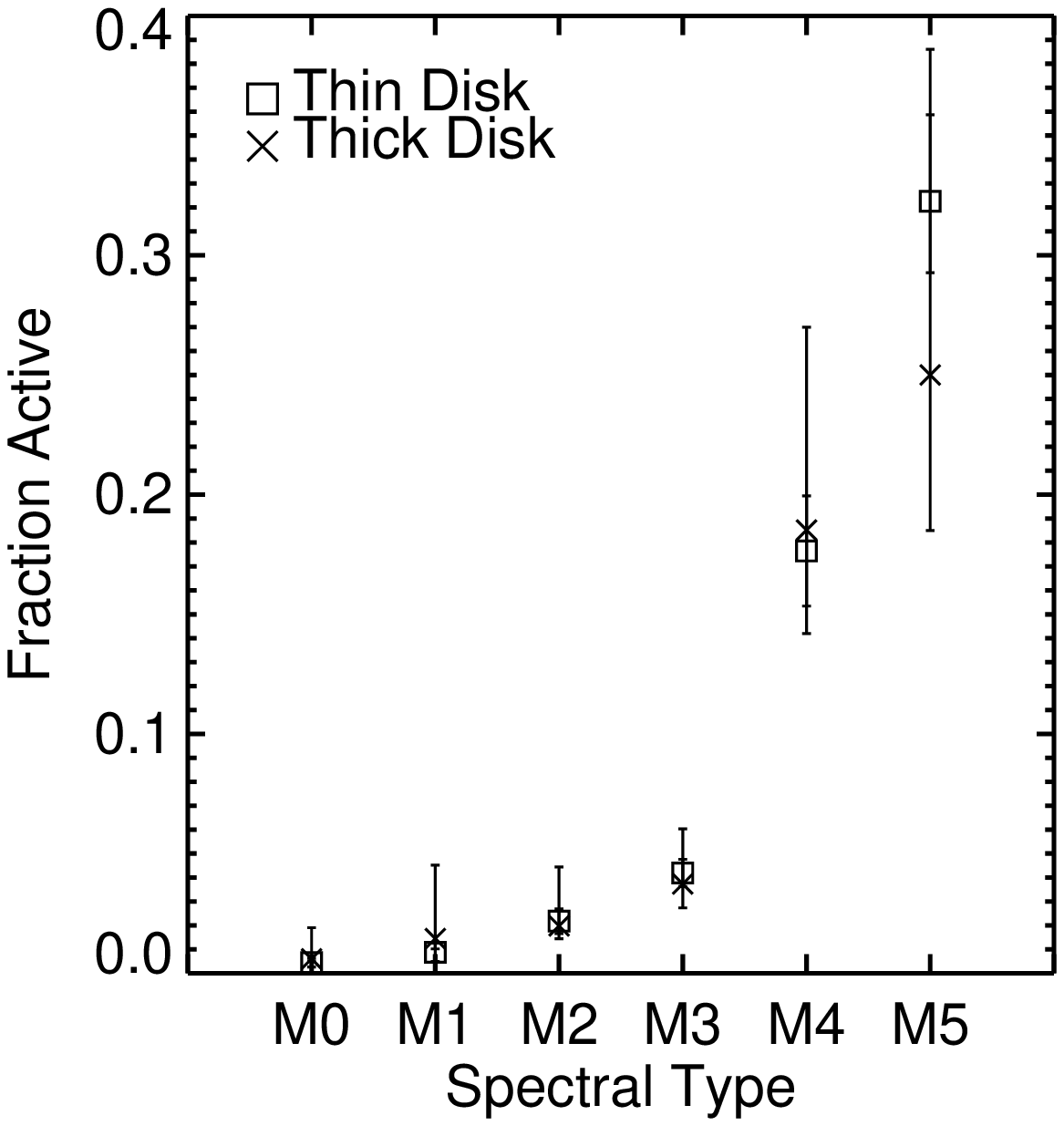}
    \caption{Shown is active fraction of stars as a function of spectral type for the thin (open squares) and thick (crosses) disk populations.  An older population would show a lower active fraction, as suggested by the thick disk stars in the M5 bin.}
\label{fig:activity}
\end{figure}

Finally, if the thick disk is an older system (as suggested by its hotter kinematics), it should possess a lower fraction of active stars \citep{2004AJ....128..426W,2006AJ....132.2507W}.  The chromospheric activity timescale varies with mass, such that higher mass stars lose their activity sooner (after $\sim$ 1 Gyr) than their low-mass counterparts ($\sim$ 10 Gyr). To observationally quantify chromospheric activity, the H$\alpha$ equivalent width (EW) is measured for each M dwarf.  We employed the technique introduced in \cite{2004AJ....128..426W} and described in \cite{2007AJ....133..531B}, selecting chromospherically active and inactive stars at each spectral type.  To be classified as active, a star must have an EW of 1 \AA, and pass additional signal-to-noise and error tests described in \cite{2004AJ....128..426W}.  Figure
\ref{fig:activity} shows the active fraction of stars as a function of spectral type for both disk populations.  While most early-type M dwarfs (M0-M3) lose their activity quite rapidly, the later types (M5) possess smaller active fractions in the thick disk, and exhibit the expected behavior of older systems.  
However, these results are only suggestive, as populations older than $\sim$ 4 Gyr require M dwarfs with types later than M5 in order to be effective probes of the age \citep{w07}. 

\section{Conclusions}\label{sec:conclusions}

The Milky Way (along with other spiral galaxies) is evidently a composite of a few major smooth components (the thin and thick disks and the halo) and many smaller structures, such as the tidal debris streams from the Sagittarius dwarf galaxy.  Using spectra, proper motions and photometry along one Galactic sight-line, we studied the kinematic and structural distributions of the smooth thin and thick disks as a function of vertical distance from the Plane.  We fit two-component Gaussians to the $UVW$ distributions as a function of height, placing new constraints on the thin and thick disk velocity dispersions.  

This sample was then employed to test the predictions of current Galactic models.  The Besan\c{c}on model was chosen for comparison, since it is widely accepted as a standard and models the kinematics of stellar populations.  We generated a suite of 25 models, and each model was sampled to be consistent with the colors and distance distribution of SLoMaSS. The bulk kinematic properties of the data and model were compared.  They agree to $\sim$ 10 km s$^{-1}$, placing a limit on the accuracy of model predictions.  However, $\sigma_W$ is poorly fit by the model, suggesting that further investigation into modeling the kinematics of the thin and thick disks is necessary.  In particular, the age-velocity dispersion plays an important role in the kinematic predictions, and a more exact definition of this relation is required.

SLoMaSS was divided kinematically, assigning membership of each star to either the thin or thick disk.  We inspected the two populations as functions of spectral type, searching for observable differences in the metallicity, colors and chromospheric activity.  While there was little observed difference between the metallicity and color distributions, the activity fraction distribution suggests that the thick disk displays an activity level consistent with an older population, although this result is purely suggestive and needs to be re-examined with later spectral types.  The lack of a strong observable signal may have several causes.  Primarily, kinematic separation of populations is not perfect, as stars with extreme kinematics in one population (i.e. the thin disk) can masquerade as members of the other population.  This would dilute observable differences, and in our case, the more numerous thin disk population may be polluting the thick disk sample, despite our best efforts to minimize this effect.  Secondly, the intrinsic properties (i.e. metallicity) of the stars that compose the thin and thick disks are drawn from overlapping distributions.  Stars that are at the edges of these distributions would also blur the distinction among observable properties, clouding the best efforts of the observer.  Further investigations using stellar tracers along the entire main sequence should alleviate some of these issues.

The authors would like to thank Neill Reid, whose comments greatly improved the scope of the analysis.  We also wish to thank Lucianne Walkowicz and Hugh Harris for their early work on this project. We also gratefully acknowledge the support of NSF grants AST02-05875 and AST06-07644 and NASA ADP grant NAG5-13111.

Funding for the SDSS and SDSS-II has been provided by the Alfred P. Sloan Foundation, the Participating Institutions, the National Science Foundation, the U.S. Department of Energy, the National Aeronautics and Space Administration, the Japanese Monbukagakusho, the Max Planck Society, and the Higher Education Funding Council for England. The SDSS Web Site is http://www.sdss.org/.

The SDSS is managed by the Astrophysical Research Consortium for the Participating Institutions. The Participating Institutions are the American Museum of Natural History, Astrophysical Institute Potsdam, University of Basel, University of Cambridge, Case Western Reserve University, University of Chicago, Drexel University, Fermilab, the Institute for Advanced Study, the Japan Participation Group, Johns Hopkins University, the Joint Institute for Nuclear Astrophysics, the Kavli Institute for Particle Astrophysics and Cosmology, the Korean Scientist Group, the Chinese Academy of Sciences (LAMOST), Los Alamos National Laboratory, the Max-Planck-Institute for Astronomy (MPIA), the Max-Planck-Institute for Astrophysics (MPA), New Mexico State University, Ohio State University, University of Pittsburgh, University of Portsmouth, Princeton University, the United States Naval Observatory, and the University of Washington. 


\end{document}